\documentclass[conference]{IEEEtran}
\usepackage{ifthen}
\usepackage[utf8]{inputenc}
\usepackage[T1]{fontenc}
\usepackage{latexsym, amssymb, dsfont}
\usepackage{amsmath,amsthm}
\usepackage{siunitx}
\usepackage{graphicx}
\usepackage{url}
\usepackage{hyperref}
\usepackage[capitalize]{cleveref}

\crefname{figure}{Fig.}{Figs.}

\usepackage{tikz}
\usepackage{circuitikz}
\usepackage{pgfplots}
\pgfplotsset{compat=1.13}
\usetikzlibrary{automata}
\usetikzlibrary{arrows,shapes.gates.logic.US,shapes.gates.logic.IEC,circuits.logic.US,calc}
\usetikzlibrary{shapes,snakes}
\usetikzlibrary{calc}
\usetikzlibrary{matrix}
\usetikzlibrary{positioning,calc}
\usetikzlibrary{through}
\usetikzlibrary{pgfplots.groupplots}

\newcommand{\dup}{\delta_\uparrow}
\newcommand{\ddo}{\delta_\downarrow}
\newcommand{\fup}{f_\uparrow}
\newcommand{\fdo}{f_\downarrow}

\newcommand{\dinfty}{\delta_{\infty}}
\newcommand{\dupinfty}{\delta^\uparrow_\infty}
\newcommand{\ddoinfty}{\delta^\downarrow_\infty}

\newcommand{\spice}{SPICE}
\newcommand{\reg}{INE}

\newcommand{\vth}{V_{th}}

\newcommand{\vdd}{V_{DD}}
\newcommand{\gnd}{\texttt{GND}}

\newcommand{\hi}{\texttt{HI}}
\newcommand{\lo}{\texttt{LO}}

\newcommand{\cin}{I}
\newcommand{\cout}{O}
\newcommand{\ior}{A}
\newcommand{\iht}{B}
\newcommand{\qi}{T}
\newcommand{\nqi}{U}

\newcommand{\boolOR}{\texttt{OR}}
\newcommand{\boolNOR}{\texttt{NOR}}
\newcommand{\orloop}{\texttt{OR Loop}}
\newcommand{\srlatch}{\texttt{SR Latch}}
\newcommand{\adder}{\texttt{Adder}}
\newcommand{\clk}{\texttt{Clock Tree}}

\newcommand{\hin}{\Delta_n^{HI}}
\newcommand{\hizero}{\Delta_0^{HI}}
\newcommand{\lon}{\Delta_n^{LO}}
\newcommand{\DI}{\Delta^I}
\newcommand{\hinum}[1]{\Delta_{#1}^{HI}}
\newcommand{\lonum}[1]{\Delta_{#1}^{LO}}

\newcommand{\invtool}{{\texttt{InvTool}}}

\definecolor{color1}{HTML}{E41A1C}
\definecolor{color2}{HTML}{377EB8}
\definecolor{color3}{HTML}{4DAF4A}
\definecolor{color4}{HTML}{984EA3}
\definecolor{color5}{HTML}{FF7F00}

\usepackage[acronym]{glossaries}
\glsdisablehyper
\setacronymstyle{long-short}
\newacronym{idm}{IDM}{Involution Delay Model}
\newacronym{ddm}{DDM}{Degradation Delay Model}

\IEEEoverridecommandlockouts 

\begin{document}
\title{Gain and Pain of a Reliable Delay Model}

\author{
  \IEEEauthorblockN{J\"urgen Maier
    \begin{minipage}[c]{1em}
      \href{https://orcid.org/0000-0002-0965-5746}{{\includegraphics[width=1em]{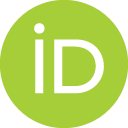}}}
    \end{minipage}
  }
  \IEEEauthorblockN{ECS Group, TU Wien, Vienna\\jmaier@ecs.tuwien.ac.at}
  \thanks{This research was funded by the Austrian Science Fund (FWF) project
    DMAC (P32431).
  }
}

\maketitle

\begin{abstract}
  State-of-the-art digital circuit design tools almost exclusively rely on pure
  and inertial delay for timing simulations. While these provide reasonable
  estimations at very low execution time in the average case, their ability to
  cover complex signal traces is limited. Research has provided the dynamic
  Involution Delay Model (IDM) as a promising alternative, which was shown (i)
  to depict reality more closely and recently (ii) to be compatible with modern
  simulation suites. In this paper we complement these encouraging results by
  experimentally exploring the behavioral coverage for more advanced circuits.

  In detail we apply the IDM to three simple circuits (a combinatorial loop, an
  SR latch and an adder), interpret the delivered results and evaluate the
  overhead in realistic settings. Comparisons to digital (inertial delay) and
  analog (\spice) simulations reveal, that the IDM delivers very fine-grained
  results, which match analog simulations very closely. Moreover, severe
  shortcomings of inertial delay become apparent in our simulations, as it fails
  to depict a range of malicious behaviors.  Overall the Involution Delay Model
  hence represents a viable upgrade to the available delay models in modern
  digital timing simulation tools.
\end{abstract}

\IEEEpeerreviewmaketitle

\section{Introduction}
\label{sec:intro}

Modern circuit designs dedicate a major share of the overall development time to
verification, in particular to answer questions like: At which pace propagate
signals through the logic? How does their appearance change on their path? Can
they be properly sampled at the output by succeeding units? To answer them
simulations are heavily utilized.

The most accurate methods currently available to determine the behavior of a
circuit are \emph{analog} simulation suites like \spice. These calculate time-
and value-continuous signal traces based on very elaborate physical models. This
is, however, a computationally expensive task such that analyses of larger
circuits quickly exceed reasonable execution times.

To reduce the overall complexity, simulations are, in general, executed in the
\emph{digital} domain, where the analog trace are abstracted by zero time
transitions between the discrete values \lo\ and \hi. Determining the temporal
evolution of digital signals throughout the circuit is done in various fashions:
The \emph{static timing analysis} (STA) solely considers the static delay of a
single input transition. Deriving reasonable values for a rising $\dupinfty$ and
falling $\ddoinfty$ transition is quite challenging, since they depend on a lot
of parameters and thus differ among gates. For this purpose extensive analog
simulations are carried out in advance, which then serve as basis for proper
predictions. Prominent examples are the \emph{Extended Current Source Model}
(ECSM) by Cadence\textsuperscript{\textregistered}~\cite{Cad:ECSM} or the
\emph{Complex Current Source Model} (CCSM) by
Synopsys\textsuperscript{\textregistered}~\cite{Syn:CCSM}.

STAs are well suited to determine, for example, the maximum clock
frequency. However, other effects, like signal degradation or interference that
may result in very short pulses, cannot be identified. For this purpose
\emph{dynamic} timing simulations, which predict time and direction of a gate's
output transitions based on time and direction of its input transitions, are
mandatory. Although several approaches are currently available (see
\cref{sec:background}), F\"ugger \textit{et al.}~\cite{FNNS19:tcad} revealed,
that solely the \gls{idm} is able to predict the behavior of a circuit solving
the short pulse filtration problem. Recently \"Ohlinger \textit{et
  al.}~\cite{OMFS20:INTEGRATION} practically applied the \gls{idm} to basic
circuits, however, primarily to evaluate the accuracy of their introduced
simulation framework. Consequently little is known about the behavioral coverage
and performance of the \gls{idm} in realistic setups.

\medskip

\textbf{Main contributions: } In this paper we are thus extending the evaluation
of the \gls{idm} to additional place \& routed circuits: For an \orloop, an
\srlatch\ and a ripple-carry \adder\ (i) analog and digital simulations are run,
(ii) the achieved results are evaluated and finally (iii) the introduced
overhead is determined. Our analyses (1) confirm the simple applicability of the
\gls{idm} stated in~\cite{OMFS20:INTEGRATION}, (2) show a high correlation
between \gls{idm} and analog simulation results and (3) reveal major
shortcomings of the approaches that are currently in use. The realistic behavioral
description, however, also leads to a significant overhead in the simulation
time of up to $250$\%.

This paper is organized as follows: In \cref{sec:background} we provide a short
introduction to existing delay models, first and foremost to the \gls{idm} and
its fundamental properties. \cref{sec:setup} then contains a description of the
simulation setup and the investigated circuits. A discussion of the achieved
\gls{idm} results and the shortcomings of inertial delay together with an
evaluation of the introduced overhead follows in \cref{sec:results}. Finally, we
conclude the paper in \cref{sec:conclusion}.


\section{Background}
\label{sec:background}

In this section we want to provide a short overview over existing delay
prediction methods, whereat we will focus in greater detail on the basics of the
\glsentryfull{idm}. For more details the interested reader is referred to the
original publication~\cite{FNNS19:tcad}.

\begin{figure}[t]
  \centering
   \includegraphics[width=0.82\columnwidth]{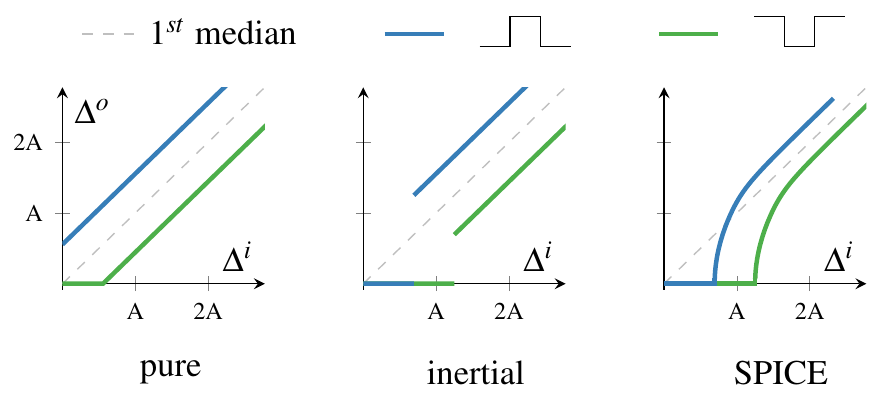}
  \caption{Output ($\Delta^o$) over input ($\Delta^i$) pulse width. Inspired
    by~\cite{MFNS19:ASYNC}.}
  \label{fig:delay_comparison}
  \vspace{-0.2cm}
\end{figure}

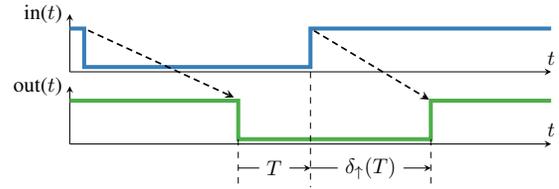
\begin{figure}[t!]
  \centering
  \scalebox{0.8}{\begin{tikzpicture}[>=stealth, scale=0.8]

	\def\lw{1.9pt}
  \draw[->] (0,0) -- (0,1.2) node [anchor=east] {in($t$)};
  \draw[->] (0,0) -- (10,0) node [anchor=south] {$t$};
  \draw[line width=\lw, color=color2] (0,0.9) -- (0.3,0.9) -- (0.3,0.1) -- (5,0.1) -- (5,0.9) -- (10,0.9);

  \draw[->] (0,-1.5) -- (0,-0.3) node [anchor=east] {out($t$)};
  \draw[->] (0,-1.5) -- (10,-1.5) node [anchor=south] {$t$};
  \draw[line width=\lw, color=color3] (0,-0.6) -- (3.5,-0.6) -- (3.5,-1.4) -- (7.5,-1.4) --
  (7.5,-0.6) -- (10,-0.6);

  \draw[dashed] (3.5,-1.4) -- (3.5,-2.4);
  \draw[dashed] (5,0) -- (5,-2.4);
  \draw[dashed] (7.5,-1.4) -- (7.5,-2.4);
  \draw[densely dashed,->,shorten >=1pt,shorten <=2pt,thick] (0.3,0.9) -- (3.45,-0.55);
  \draw[densely dashed,->,shorten >=1pt,shorten <=2pt,thick] (5,0.9) -- (7.5,-0.6);

  \node (T) at (4.25,-2) {$T$};
  \node (DT) at (6.25,-2) {$\dup (T)$};
  \draw[-] (T) -- (3.5,-2);
  \draw[->] (T) -- (5,-2);

  \draw[->] (DT) -- (7.5,-2);
  \draw[-] (DT) -- (5,-2);
\end{tikzpicture}}
  \caption{The delay value $\dup$ as function of $T$. Taken
    from~\cite{OMFS20:INTEGRATION}.}
  \label{fig:single_history}
\end{figure}

The most straight-forward approach for digital delay estimation is clearly the
\textit{pure} delay which introduces a retardation of $\dupinfty$
resp. $\ddoinfty$. Obviously this leads to a linear relationship between input
($\Delta^i$) and output ($\Delta^o$) pulse width, i.e., the time difference
between transitions of opposite directions.  \cref{fig:delay_comparison} shows
the relationships for up-pulses (falling after rising transition) and
down-pulses (falling before rising). The also depicted \emph{inertial} delay is
very similar, with the difference that pulses below a certain threshold are not
propagated. Although a comparable behavior is observable in analog \spice\
simulations, they mainly differ by a gradual degradation of
$\Delta^o$. Capturing this effect is mandatory to realistically model the
generation of glitches, i.e., very short pulses. Juan-Chico \textit{et
  al.}~\cite{IDDM} showed that inertial delay is unfit for this task which lead
to the development of the \gls{ddm}~\cite{BJV:DDM05}. This approach
predicts the delay value using nonconstant delay functions $\dup(T)$
resp. $\ddo(T)$. The parameter $T$ is thereby defined as the time span from the
last output transition to the current input transition, as is shown in
\cref{fig:single_history}.

For long input pulses (large values of $T$), $\delta(T)\approx \dinfty$ can be
assumed to be constant, comparable to inertial/pure delay. Decreasing the pulse
width eventually leads to a decline of $\delta(T)$,
cp. \cref{fig:delay_comparison}, and hence to significant degradation. Finally,
the equilibrium point $-T=\delta(T)$ is reached, which generates a zero-time
glitch at the output. For even shorter input pulses the model schedules the
digital output transitions in the wrong temporal order, e.g., when starting at
\lo\ a falling before a rising transition. In this case we speak of
\emph{cancellation} and both transition are removed. In the analog domain this
corresponds to sub-threshold trajectories.

Despite its removal during cancellation, the latest transition time is still
crucial, as it serves as reference point for the calculation of the succeeding
$T$.  Since there are no more threshold crossings at the output available, that
could be used to determine the delay, $\delta(.)$ has to be predicted. For this
purpose \gls{ddm} simply extends the fitting of $\delta(T)$ derived for
$T>-\delta(T)$. While this seems, at a first glance, like a legitimate choice it
causes the delay estimation to fail in certain circumstances. In detail F\"ugger
\textit{et al.}~\cite{FNS13} were able to prove that all existing approaches,
including \gls{ddm}, pure and inertial delay, cannot faithfully model a circuit
that solves the short-pulse filtration problem (SPF). In consequence the
\glsdesc{idm}~\cite{FNNS19:tcad} was developed, with the distinguishing property
that input pulses with $\Delta_i \rightarrow 0$ have diminishing effect on the
output.  An interpretation of the model in the analog domain is described by the
authors in the following fashion: The digital input signal first passes a pure
delay component and is then transformed to the analog domain using two unique
waveforms ($\fup$ from \lo\ to \hi, $\fdo$ from \hi\ to \lo), which are switched
instantaneously upon a transition.  Finally the analog trajectory is fed into a
comparator, which issues a digital output event whenever the threshold voltage
$\vth$ is crossed.

Although \gls{ddm} and especially \gls{idm} have much higher expressive power
modern circuit designs still heavily rely on the simple pure and inertial delay
models. This is not surprising, given the very good integration in
state-of-the-art simulation suites and thus its simple applicability:
Implementations based on popular hardware description languages like VHDL
Vital~\cite{IEEE:VITAL} or Verilog~\cite{IEEE:Verilog} are widespread.  Albeit
there is a distinguished simulation tool for \gls{ddm}~\cite{HALOTIS} available,
it is shipped as a separate executable, making it demanding to integrate the
simulation into an existing design flow. To circumvent this problem for the
\gls{idm}, \"Ohlinger \textit{et al.}~\cite{OMFS20:INTEGRATION} developed
the~\invtool, whose VHDL procedures simply have to be linked and thus enforce no
changes on the established work flow.


\section{Experimental setup}
\label{sec:setup}

To determine the behavioral coverage and performance of the \gls{idm} we chose
to run analog and digital simulations. The respective framework, which utilizes
the \SI{15}{\nm} Nangate Open Cell Library with FreePDK15$^\text{TM}$ FinFET
models~\cite{Nangate15} ($\vdd=\SI{0.8}{\V}$), is described in the sequel.  A
presentation and evaluation of the results follows in \cref{sec:results}. The
simulation data is freely available
on-line\footnote{\url{https://github.com/jmaier0/idm_evaluation}}.

\subsection{Design Flow}

For the sake of realistic results we utilize the
Cadence\textsuperscript{\textregistered} tools Genus\textsuperscript{TM} and
Innovus\textsuperscript{TM} (version 19.11) to place \& route the design and
automatically extract the parasitics (.spef format) and static delay values
(.sdf format). For analog transient simulations we back-annotate the extracted
parasitics to a transistor level model, which is then executed using
Cadence\textsuperscript{\textregistered}
Spectre\textsuperscript{\textregistered} (version 19.1). Note that these results
serve as golden reference for the digital predictions, which enables a quick and
simple evaluation regarding the correctness and behavioral coverage.

The digital simulations are run with Mentor\textsuperscript{\textregistered}
ModelSim\textsuperscript{\textregistered} (version 10.5c), which reads the .sdf
file to parameterize the circuit netlist generated by
Innovus\textsuperscript{TM}.  Two prediction approaches were executed: The
default one provided by the tool (\reg), essentially an inertial delay, and the
\glsdesc{idm} (\gls{idm}). For the latter we used the
\invtool\footnote{\url{https://github.com/oehlinscher/InvolutionTool}}, to
retrieve the desired $\exp$-channel model (\emph{ea/p\_exp\_channel.vhd}).
Note that the \texttt{XOR} gate could not be generated automatically and had to
be created manually. Since the .sdf file just defines the static delays
$\dinfty^{\uparrow/\downarrow}$, we set, for the sake of simplicity, the pure
delay to a constant value of $\SI{1}{\ps}$.

We want to emphasize at this point that we were able to confirm the simplicity
of applying \gls{idm} to an existing design flow. Starting from the test setup
for \reg\ we solely had to compile and link the respective \gls{idm} files.
Nevertheless, we were not able to reuse our testbench since the \gls{idm} and
the tool's delay model are implemented in differing hardware description
languages (Verilog vs. VHDL). Some commands, such as forcing signals, do not
properly work across language boundaries, which made it necessary to duplicate
the testbench while, of course, conserving the same behavior.

\subsection{Circuits}

In the sequel we introduce the circuits used in our simulations. Note that
additional buffers, which we added at the in- and output to emulate the settings
far away from the chip boundaries, are not shown.

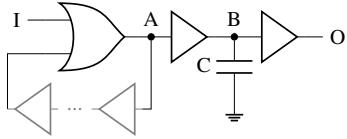
\begin{figure}[ht!]
  \centering
  \scalebox{0.8}{\begin{circuitikz}[
	conn/.style={circle, minimum width=3pt, inner sep=0, fill=black, draw},
	opt/.style={color=gray, draw},
]
	\def\sc{0.6}

	\draw
		(0,0) node[american or port, draw] (OR) {}
		(1,0) node[buffer, scale=\sc, draw] (HT) {}
		(2.5,0) node[buffer, scale=\sc, draw] (O) {}
		(-0.2,-1.2) node[buffer, scale=\sc, opt, xscale=-1] (FB1) {}
		(-0.9,-1.2) node[opt, draw=none] {...}
		(-1.6,-1.2) node[buffer, scale=\sc, opt, xscale=-1] (FB2) {}
		(OR.out) -- (HT.in) node[pos=0.5, conn, label={90:\ior}] (CONN) {} 
		(HT.out) -- (O.in) node[pos=0.5, conn, label={90:\iht}] (CONNC) {}
		(O.out) -- ++(0.3,0) node[right] {\cout}
		(CONNC) to [/tikz/circuitikz/bipoles/length=1cm,C, l_=C] ++(0,-1) node[ground, scale=0.7] {}
		(CONN) |- (FB1.in)
		(FB2.out) |- (OR.in 2)
		(OR.in 1) -- ++(-0.3,0) node[left] {\cin}
	;

\end{circuitikz}}
  \caption{\orloop\ gate level implementation.}
  \label{fig:or_loop}
\end{figure}

\subsubsection{\orloop}
\label{sec:setup_or_loop}

In its bare form the circuit shown in \cref{fig:or_loop} has been used
in~\cite{FNNS19:tcad} for proofing the faithfulness of \gls{idm} regarding the
SPF problem. It utilizes an arbitrary amount of buffers to create a
combinatorial loop, whereat up-pulses are coupled in via a single
\boolOR-gate. Based on the input pulse width $\DI$ the signal may oscillate for a
possibly infinite amount of time before vanishing or setting the loop to
\hi. Depending on the length of the feedback path either distinguished pulses or
intermediate voltage values are observable. While the former corresponds to a
simple ring oscillator the latter depicts \emph{metastability}~\cite{G11:DTC},
which leads to a wide range of problematic behaviors such as late output
transitions or a spurious mapping to \lo\ and \hi\ among succeeding gates. Thus
it is crucial to model metastable upsets in a suitable fashion in the digital
domain. To ease the descriptions of oscillations and metastability we are going
to use $\hin$ and $\lon$ to denote the high respectively low time of the
$n^{\text{th}}$ oscillation at node \ior.

To focus on different characteristics we ran simulations with a varying number
of buffers in the feedback path, which effectively varies the loop delay. We are
aware that the same could also be achieved by adding capacitances, however, in
our setup this would lead to significantly different results since a capacitance
serves as a low pass filter and thus suppresses short, high frequency, pulses
very effectively. Using multiple buffers in succession, on the contrary,
conserves the signal shape such that oscillating signals can be
generated. Nonetheless we artificially added a large capacitance at node \iht,
which enables us (i) to analyze the internal behavior and (ii) to reveal
shortcomings of the delay models more easily.

\begin{figure}[ht!]
  \centering
  \scalebox{0.8}{\begin{circuitikz}[
	conn/.style={circle, minimum width=3pt, inner sep=0, fill=black, draw},
	opt/.style={color=gray, draw},
]
	\def\sc{0.6}

	\draw
		(0,0) node[american nor port, draw] (OR1) {}
		(1,0) node[buffer, scale=\sc, draw] (BNQ) {}
		(0,-2) node[american nor port, draw] (OR2) {}
		(1,-2) node[buffer, scale=\sc, draw] (BQ) {}
		(OR2.in 1) node [anchor=out, buffer, scale=\sc/2, draw] (BQI) {}
		(OR1.in 2) node [anchor=out, buffer, scale=\sc/2, draw] (BNQI) {}
		(OR1.out) node[conn, label=90:\nqi] {} -- ++(0,-0.5) -- ($(BQI.in)+(-0.1,0.5)$) |- (BQI.in)
		(OR2.out) node[conn, label=-90:\qi] {} -- ++(0,+0.5) -- ($(BNQI.in)+(-0.1,-0.5)$) |- (BNQI.in)
		(OR1.in 1) ++(-0.2,0) node [left] {$S$}
		(OR2.in 2) -- ++(-0.2,0) node [left] {$R$}
		(OR1.out) -- (BNQ.in)
		(BNQ.out) -- ++(0.3,0) node[right] {$\overline Q$}
		(OR2.out) -- (BQ.in)
		(BQ.out) -- ++(0.3,0) node[right] {$Q$}
	;

\end{circuitikz}}
  \caption{\srlatch\ gate level implementation.}
  \label{fig:sr_latch}
\end{figure}
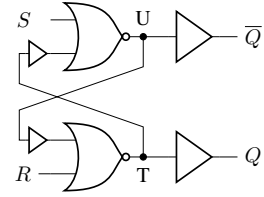

\subsubsection{\srlatch}
\label{sec:setup_latch}

The second circuit we chose is the \srlatch\ as shown in \cref{fig:sr_latch}, a
well-known circuit with the possibility for metastability and slightly improved
complexity. Note that we added a single buffer on the coupling paths between the
\boolNOR-gates, to pronounce the observable effects and thus ease their
detection.

The Set Reset Latch operates very intuitively: If the set ($S$) input turns \hi,
$Q$ switches to \hi, for a \hi\ on the reset ($R$) input, $Q$ changes to \lo.
$\overline{Q}$ represents the inverse of $Q$ and thus shows the opposite
behavior. Both inputs set to \hi\ leads to an intermediate voltage value at
nodes \nqi\ \& \qi\ and thus has to be prevented.  Note the similarities between
\srlatch\ and \orloop: If one input is \lo, the \srlatch\ behaves, w.r.t. the
other one, just like the \orloop: Very short pulses are blocked, very long ones
immediately set the loop, while ones in between may lead to
metastability. Significantly different behavior is possible, however, if both
inputs are allowed to change. While one steers the loop into a metastable state
the other one can either support or impair its resolution, a fact that we will
exploit in our simulations.

\begin{figure}[ht!]
  \centering
  \scalebox{0.8}{\begin{tikzpicture}[
	fa/.style={rectangle, minimum width=3em, minimum height=3em, inner sep=0, draw},
]
	\def\x{1.8};	
	\def\y{-0.3};	

	\node[fa]	 (FA1) at (0,0) {FA};
	\node[fa]	 (FA2) at (\x,0) {FA};
	\node[fa]	 (FAN) at (2.5*\x,0) {FA};
	\node at ($(FA2)!0.5!(FAN)$) {$\hdots$};

	\draw
		(FA1.240) -- ++(0,\y) node[below] {$A_0$}
		(FA1.300) -- ++(0,\y) node[below] {$B_0$}
		(FA1.90) -- ++(0,-\y) node[above] {$S_0$}

		(FA2.240) -- ++(0,\y) node[below] {$A_1$}
		(FA2.300) -- ++(0,\y) node[below] {$B_1$}
		(FA2.90) -- ++(0,-\y) node[above] {$S_1$}
		(FA2.east) --  ++(-\y,0) node[below] {$C_2$} 

		(FAN.240) -- ++(0,\y) node[below] {$A_n$}
		(FAN.300) -- ++(0,\y) node[below] {$B_n$}
		(FAN.90) -- ++(0,-\y) node[above] {$S_n$}
		(FAN.west) --  ++(\y,0) node[below] {$C_n$} 
		(FAN.east) --  ++(-\y,0) node[right] {$S_{n+1}$} 

		(FA1) -- node[below] {$C_1$} (FA2) 
		(FA1.west) --  ++(\y,0) node[left] {$0$} 
	;

\end{tikzpicture}}
  \caption{\adder\ gate level implementation.}
  \label{fig:adder}
\end{figure}
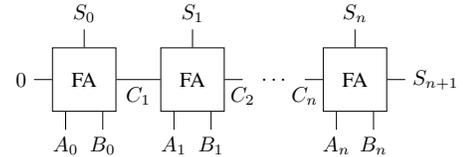

\begin{figure*}[tb]
  \centering
  \includegraphics[width=0.9\linewidth]{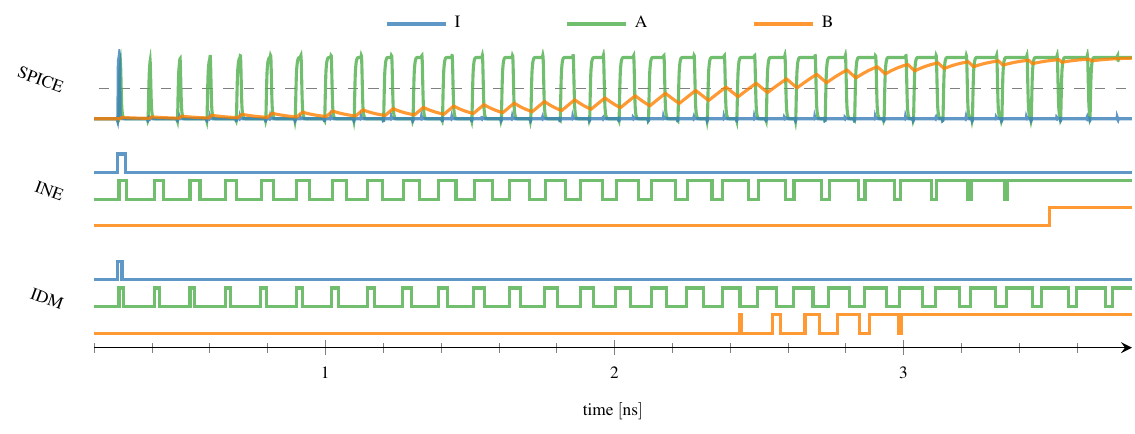}
  \caption{Analog and digital simulation results for the \orloop\ with long feedback.}
  \label{fig:or_loop_longfb}
\end{figure*}

\subsubsection{\adder}
\label{sec:setup_adder}

To investigate the scaling of the \gls{idm} and its predictions on loop-free
circuits we also simulated a simple ripple carry adder as shown in
\cref{fig:adder}, whereat we used $n=4$. Each full adder block FA is defined on
the gate level and implements the equations
\begin{align*}
  S_i &= C_i \oplus A_i \oplus B_i\\
  C_{i+1} &= (C_i \land (A_i \oplus B_i)) \lor (A_i \land B_i)
\end{align*}

Out of the manifold input possibilities, those leading to a maximum number of
transitions are the most interesting for our analysis, as they allow an
investigation of the whole circuit in a single simulation run. For this purpose
we chose $B_0B_1B_2B_3=1111$, $A_0A_1A_2A_3=0000$ and introduced an up-pulse on
signal $A_0$. For a down-pulse on signal $A_0$ we used a very similar setup,
with the sole difference of setting $A_0A_1A_2A_3=1000$ initially.


\section{Results}
\label{sec:results}

In this section we present and compare the analog resp. digital simulation
results for the circuits introduced in \cref{sec:setup}.  We start by studying
oscillatory behavior and its digital counterpart for the \orloop\ with long
feedback delay. Subsequently we remove the buffers from the feedback path and
investigate the effects on the (significantly changing) analog and (only
slightly differing) digital simulation results. Afterwards we use the \srlatch\ to
demonstrate the superior modeling power of the \gls{idm}, which, in contrast to
inertial delay, predicts the metastable behavior quite well. Simulations
of the \adder\ confirm the superiority of the \gls{idm} but also reveal
inaccuracies. Finally we present an extensive comparison of the overhead and
thus the price that has to be paid.

\subsection{\orloop\ with Long Feedback}
\label{sec:or_loop_longfb}

For our first experiments we added thirty buffers to the feedback path. In this
setup it is, for $\dupinfty = \ddoinfty$, possible to generate multiple periodic
signals in the loop since the signal rise/fall time is significantly smaller
than the overall delay of the loop. However, the static delay values extracted
after place \& route did not match: Rising transitions are delayed less than
falling ones, leaving exactly one $\DI$ that perfectly compensates the increase
in $\hin$ by pulse degradation effects and thus creates infinite oscillation.

\subsubsection{\spice}

\cref{fig:or_loop_longfb} (top) shows the analog simulation results for an
initially very short pulse that grows and eventually settles the loop at
$\vdd$. Clearly visible is the impact of the high capacitive load. Since the
transitions at node \ior\ are very quick compared to node \iht\ it actually
seems as if charging and discharging curves are switched immediately when an
input transition occurs\footnote{Recall that this perfectly matches the analog
  domain model of the \gls{idm}}. Consequently the threshold (dashed line) is
crossed multiple times, whereat the time difference between rising and falling
trajectories strictly increases.

Noteworthy is the high sensitivity of the feedback loop in this operation region
and thus also the very low probability to reach such a state. We had to vary
$\DI$ in steps of \SI{1}{\as} in order to eventually generate an oscillation
trace inside the loop that lasted at most \SI{4}{\ns}.

\subsubsection{\reg}

At a first glance the inertial delay results shown in \cref{fig:or_loop_longfb}
(middle) look comparable. The short initial pulse increases until the loop is
constant \hi\ and thus also node \iht\ gets \hi. However, on closer examination
a severe shortcoming is revealed: The shown pulse is the shortest one that can
be inserted into the loop, as smaller ones are removed by a high-delay buffer at
the input. This indicates a general problem: A gate with long delay at the front
may remove a big share of the input pulses, which can include highly relevant
ones. Consequently it is impossible to detect any infinite or decaying
oscillations for the shown circuit using \reg.

Note that the rising transition at node \iht\ only occurs after the loop has
fully settled, i.e., the oscillations have ceased. This can again be explained
by the big delay of the succeeding gate, which thus serves as a metastability
filter. This does, however, not correspond well to the analog simulations, where
the threshold is already crossed way before the loop is fully locked. Therefore
\reg\ is not suited to properly describe the exact behavior of the circuit
in such circumstances. In particular, it is impossible to achieve pulses at node
\iht\ for the inertial delay model: only a single transition is observed or none
at all.

\subsubsection{\glsentrytext{idm}}

Compared to \reg\ the \glsentrylong{idm} achieves a much more fine grained
behavioral description. First and foremost, any value of $\hizero$ can be
generated, also ones that quickly decay. \cref{fig:or_loop_longfb} (bottom)
shows a simulation with increasing $\hin$ for ascending $n$: Considering the
model representation in the analog domain presented in \cref{sec:background}
this corresponds to utilizing $\fup$ for an increasing and $\fdo$ for a
decreasing amount. Consequently the mean analog value steadily rises, eventually
crossing $\vth$ and resulting in the digital oscillations on node \iht\ shown in
the figure.

By properly tuning $\DI$ it is even possible to achieve an infinite pulse train,
i.e., one that perfectly recreates itself. Note that, although the loop is highly
unstable in this configuration, not a single transition on \iht\ could be
observed, which reveals a problem of the \gls{idm}: Depending on the value of
the discretization threshold voltage $\vth$, zero, one or infinitely many
transitions are indicated for the same analog trajectory.

\subsubsection{Comparison}

\begin{figure}[tb]
  \centering
  \includegraphics[width=0.9\linewidth]{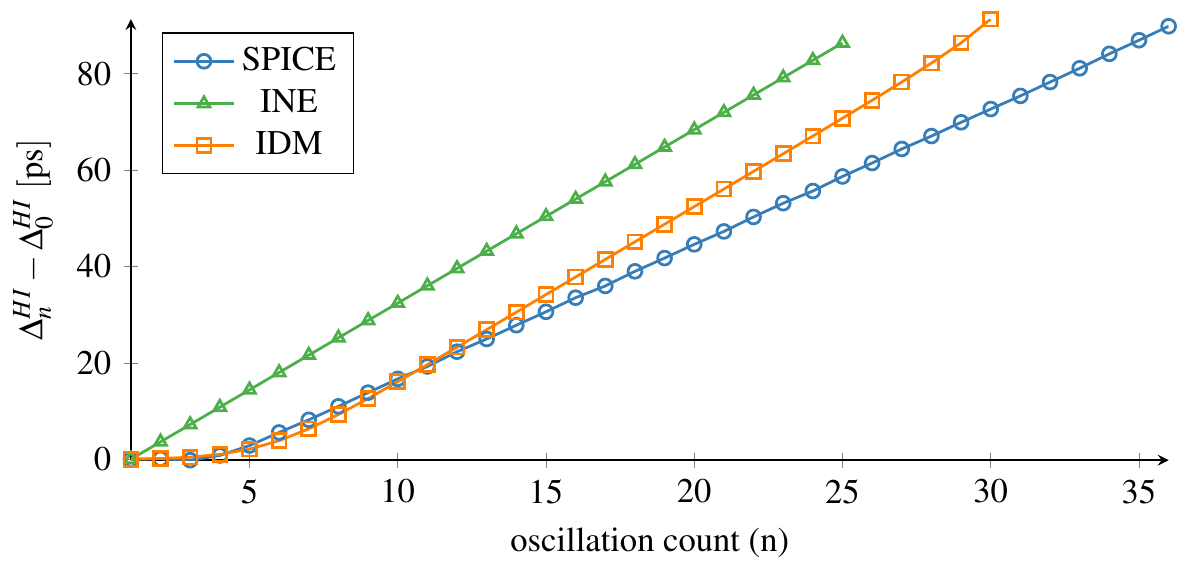}
  \caption{Increase of the pulse train high time compared to its initial
    value.}
  \label{fig:pw_increase}
\end{figure}

\gls{idm} and \reg\ also handle the evolution of $\hin$, shown in
\cref{fig:pw_increase}, differently. Consider that the rate of growth is
determined as the gap between falling and rising transition delay, which is
constant for \reg\ due to fixed delay values. Consequently a linear increase of
$\hin$ can be observed.  \spice\ and \gls{idm}, however, show a quite different
behavior: For small $n$, $\hin$ increases only marginally, as the circuit
initially operates near the metastable point, i.e., where pulses recreate
themselves. With increasing pulse width the rate, however, quickly ramps up.

Very interesting is the nonlinear increase of the \gls{idm}. Intuitively, $\hin$
is expected to settle at a constant rate, since for large values of $T$ the
\gls{idm} and inertial delay are equal. While this is true, one has to consider
that the increase in $\hin$ causes a drop of $\lon$, which eventually
experiences pulse width degradation and thus further enhances the increase rate
of $\hin$.

\subsection{\orloop\ with Direct Feedback}
\label{sec:or_loop_nofb}

Recall that the buffer count in the feedback path is a very sensitive parameter
with major implications: Reducing the delay of the loop moves rising and falling
transitions closer together, while leaving the rise and fall time
untouched. Eventually the single transitions will merge meaning that \gnd/$\vdd$
are not reached any more.

The effects on the infinite oscillatory behavior are as follows: As long as
there is at least one gate in the loop still performing full range switching,
which is possible due to differing parasitics, oscillations with a reduced
amplitude, i.e., within the range $[V_L, V_H]$ with $V_L > \gnd$ and $V_H<\vdd$,
are possible (cf. \cref{sec:latch}).  Due to the lower amplitude, the time
between succeeding threshold crossings declines and thus the oscillation
frequency increases. Further reducing the delay finally leads to a damped
oscillation, whereat the damping factor is increased with decreasing delay. For
the simulations presented in the sequel we removed all gates and thus force a
direct transition to the constant metastable voltage.

\subsubsection{\spice}

\begin{figure}[tb]
  \centering
  \includegraphics[width=0.8\linewidth]{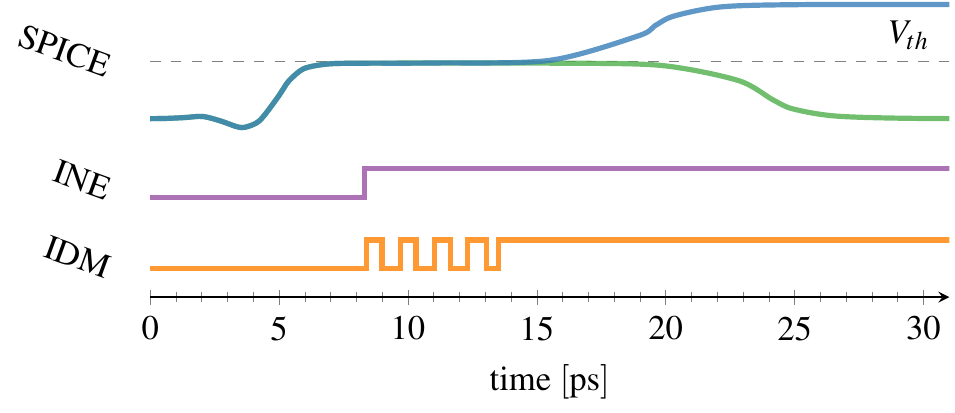}
  \caption{Analog and digital simulation results of node \ior\ for the \orloop\ with
    direct feedback path.  }
  \label{fig:or_loop_nofb}
\end{figure}

Analog simulations confirm our intuitive explanation. \cref{fig:or_loop_nofb}
shows two traces on node \ior, which stay at a constant value near $\vth$ for
some time and then resolve to \lo\ in one case and to \hi\ in the other one. The
facts that the corresponding $\DI$ only differ by \SI{1}{\as} and, nonetheless,
it is only possible to stay in the metastable state for a few picoseconds,
indicate the very high sensitivity of this circuit configuration.

\subsubsection{\reg}

As described in \cref{sec:or_loop_longfb} the shaping gates at the input filter
many incoming pulses. In fact, only those longer than the delay of the storage
loop are able to pass, causing an immediate switch to \hi. Consequently, for
\reg, the simulation either delivers a single rising transition on all wires or
none at all. While this might seem reasonable at a first glance, the metastable
state, and thus the increase in delay, are not revealed, suggesting falsely a
settled and well defined behavior.

\subsubsection{\glsentrytext{idm}}

Although the analog simulations did not show any $\vth$ crossing during
metastability, the \gls{idm} again delivers an oscillatory behavior, which seems
to be awfully wrong. Considering, however, the analog representation, more
specifically the switching between $\fup$ and $\fdo$, it becomes apparent that
the closest the analog trajectory in the \gls{idm} can get to a constant
intermediate value is to oscillate around it. Therefore a pulse train is used to
indicate metastability.

The fact that the \gls{idm} used a pulse train to describe both real
oscillations in \cref{sec:or_loop_longfb} as well as metastability begs the
question: How can these scenarios be distinguished? The answer is
discouraging. Solely based on the digital predictions this is impossible. The
major difference among oscillatory traces are $\hin$ respectively $\lon$, which,
however, do not yield much information on their own. Only in combination with
the switching waveforms $\fup$ \& $\fdo$ or the static delays
$\dinfty^{\uparrow/\downarrow}$ it is possible to estimate the voltage gain
during the \hi\ resp. \lo\ period. As s rule of thumb one has to expect damping
if $\hin$ ($\lon$) is approximately or lower than $\dinfty^{\uparrow}$
($\dinfty^{\downarrow}$).

In our setup we extracted $\dupinfty=\SI{4.6}{\ps}$ and
$\ddoinfty=\SI{5.8}{\ps}$ for the \boolOR-gate, which is clearly more than
$\hin$ respectively $\lon$ in \cref{fig:or_loop_nofb}. Although this seems very
disadvantageous for the \gls{idm} be advised that also for \reg\ comparisons
with the delay values are necessary to determine if a pulse is close to
suppression.  Since knowing the peak values in the analog domain is so important
we are currently working on an extension of the \gls{idm} that is able to
indicate the underlying analog waveform. This is, however, only suited for
rough estimations and is not intended to replace analog simulations.

Overall it has to be stated that an oscillating simulation trace does not
automatically indicate an undesired behavior. Just as periods reach a circuit
dependent range ill shaped pulses or even metastability have to be inferred.

\subsection{\srlatch}
\label{sec:latch}


\begin{figure}[tb]
  \centering
  \includegraphics[width=0.97\linewidth]{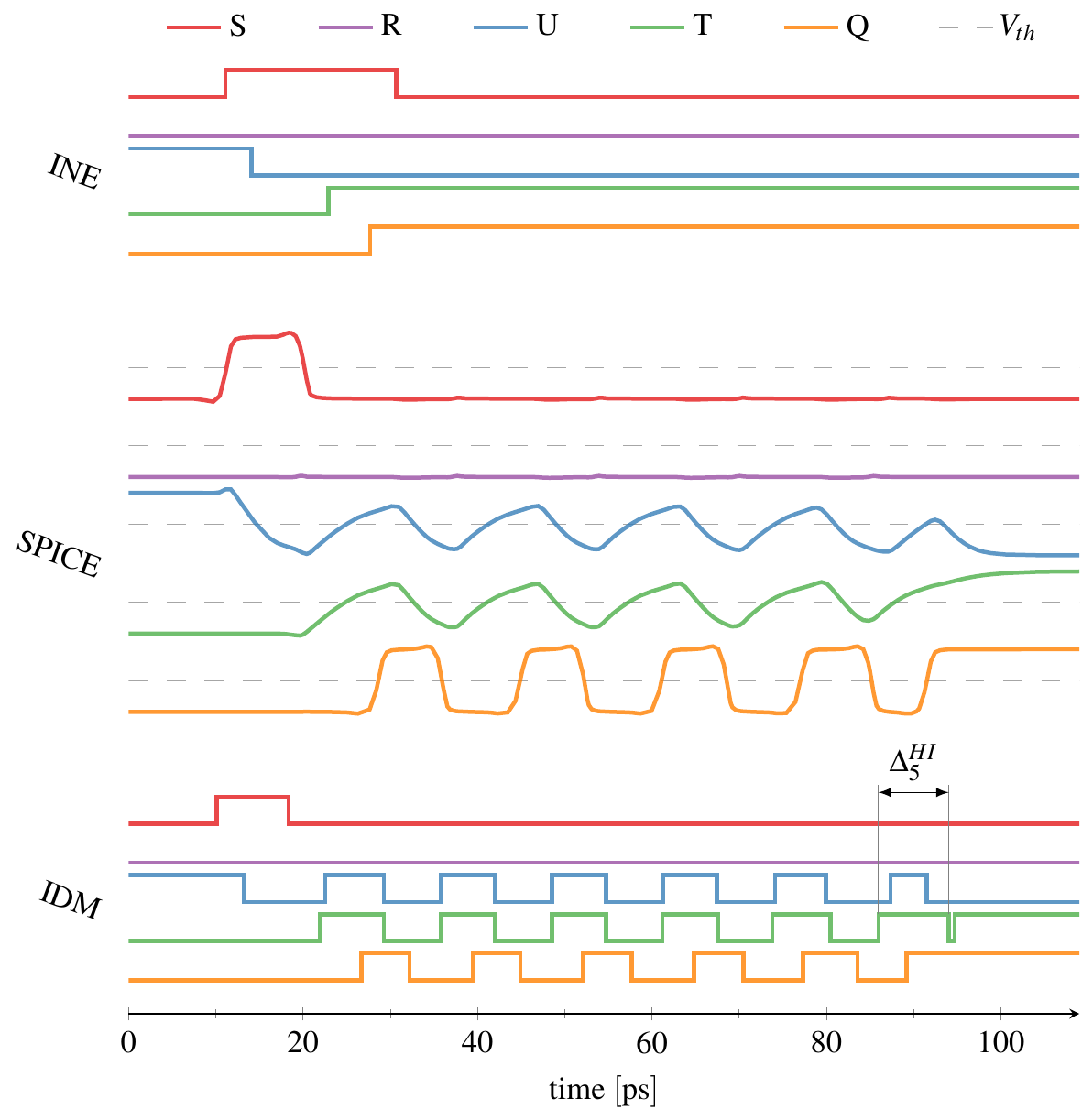}
  \caption{Simulation results showing metastability in the \srlatch.}
  \label{fig:sr_latch_set}
\end{figure}

After studying the general behavior of digital simulation approaches on the
rather synthetic \orloop, we now turn to the common \srlatch. Interestingly,
\reg\ again fails to cover very important parts of the real behavior and thus
delivers overly optimistic results, while the \gls{idm} stays close to the
analog trace. The latter even enables us to explore unfavorable input
conditions, which we will use to artificially prolong metastability.

\subsubsection{Set or Reset Input Pulse}

Since setting either $S$ or $R$ \lo\ degrades the \srlatch\ to the \orloop\
w.r.t. the other input, simulations shown in \cref{fig:sr_latch_set} lead to
similar results. For \reg\ the shortest pulse, that is able to pass the input
buffers, once again immediately sets the loop, leading to a single output
transition. This strongly contradicts the analog simulation, which shows (non
full-range) oscillations on all wires. As discussed earlier, such a behavior is
possible if one of the gates in the path, in this case the buffers, still issue
full range waveforms.

In contrast, the \gls{idm} describes the behavior in, and also the resolution
out of, metastability faithfully, which enabled us to search for ``malicious''
input conditions that prolong the metastable state. In \cref{fig:sr_latch_set} a
very long \hi\ phase ($\hinum{5}$) on node \qi\ is visible as Q switches to
constant \hi. To prevent the oscillation from resolving it would be necessary to
decrease $\hinum{5}$ and simultaneously increase $\lonum{5}$. Reiher \textit{et
  al.}~\cite{RGJ18:ASYNC} described a similar effect when ``kicking''
synchronizers, i.e., abruptly changing an internal voltage value, which also led
to a potential extension of the metastable state.

It can be easily retraced that a properly placed up-pulse on the reset input $R$
does the trick. Essential for success is the time of the rising transition, as
it determines the width of $\hinum{5}$. On the contrary the falling transition
can be issued at any point in time during the \hi\ period of the other
\boolNOR-gate input, since in this case the reset input is masked anyway.

\subsubsection{Set and Reset Input Pulse}

\begin{figure}[tb]
  \centering
  \includegraphics[width=0.97\linewidth]{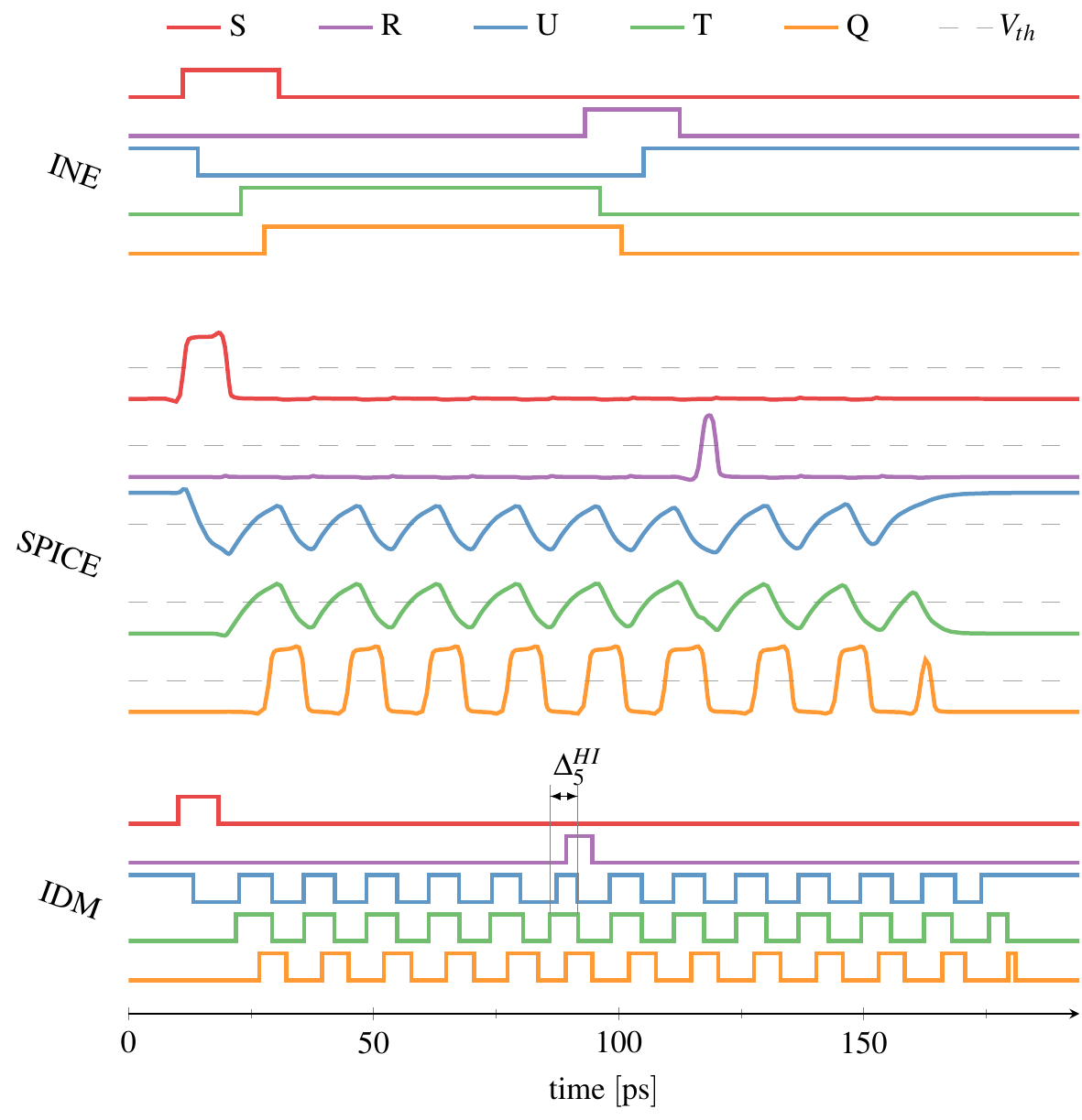}
  \caption{Simulation results of steering the \srlatch\ back into metastability.}
  \label{fig:sr_latch_set_reset}
\end{figure}

To verify our predictions, we extended the previous simulation by a pulse on
input $R$. Results for \reg, shown in \cref{fig:sr_latch_set_reset} (top),
reveal solely one additional output transition and thus suggest a fully settled
circuit, while completely hiding the potential instabilities.

On the contrary, \spice\ simulations shown in \cref{fig:sr_latch_set_reset}
(middle) confirm our predictions. Not only is metastability extended but also a
resolution to \hi\ is forced.  We want to emphasize that the signal on $R$ used
to prolong metastability is too short to have any impact on a fully settled
memory loop, which was revealed by further simulations. Only in combination with
this particular unstable circuit state a change in value becomes
possible. Consequently, a close observation of unstable states and short pulses
is very important.


Finally an execution of the \gls{idm} shown in
\cref{fig:sr_latch_set_reset} (bottom) delivers exactly the predicted behavior.
Cutting $\hinum{5}$ indeed sets the loop back into metastability, resulting in a
very realistic representation of the underlying analog behavior.


\subsection{\adder}
\label{sec:adder}

\begin{figure}[tb]
  \centering
  \includegraphics[width=0.97\linewidth]{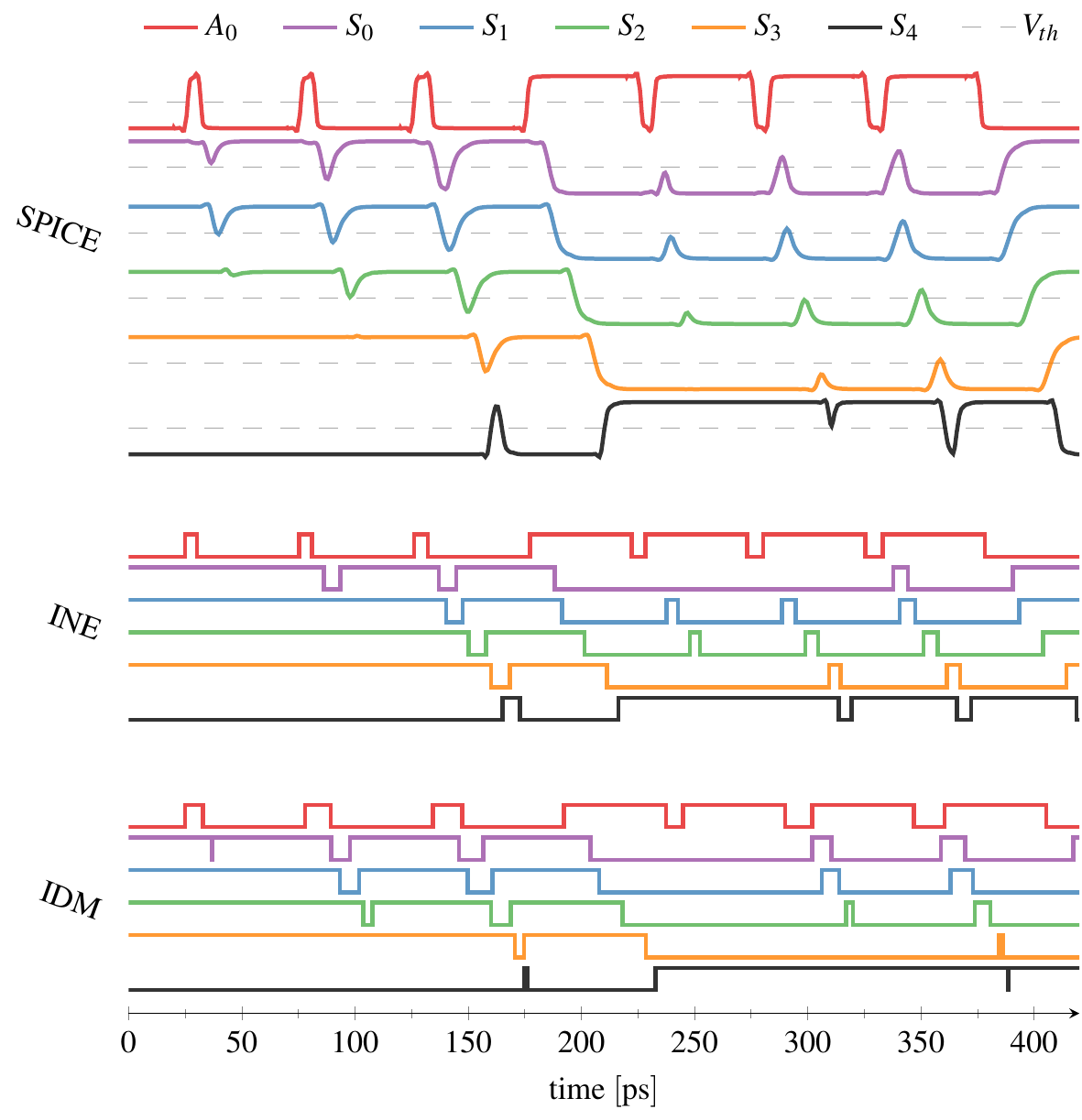}
  \caption{Simulation results of the \adder\ with an input glitch.}
  \label{fig:adder_trans}
\end{figure}

At last we turn to the four Bit \adder\ to evaluate the scaling potential of the
\gls{idm}.  Analog \spice\ simulation results [see \cref{fig:adder_trans} (top)]
clearly show the propagation of the input pulse through the unit and the
corresponding degradation. Whether a pulse is observed on output $S_i$ depends
on (i) the initial input pulse width on signal $A_0$ and (ii) the path length
from $S_i$ to $A_0$. The longer the path the bigger the input signal has to be
to still have an impact. Interestingly the carry signals $C_{i+1}$ seem to be
faster than the ones representing the corresponding sum value $S_i$, which can
be seen very clearly by comparing $S_3$ and $S_4$ (the latter is actually the
carry signal of the last full adder). While $S_3$ still barely crosses the
threshold $S_4$ already reaches all the way to \gnd/$\vdd$.

Overall these results show the threat caused by glitches: Due to the differing
path lengths through the circuit the input signal generates a varying number of
output pulses with decreasing pulse widths on signals $S_i$ which elevate the
chance to violate the setup and hold times of succeeding flip-flops. Furthermore
we want to emphasize that in this circuit a metastable input value has the
chance to spread to five output signals and thus the effect of a single upset
gets multiplied. This shows, once more, the importance of faithfully predicting
glitches and metastability in the first place.

For \reg\ a very inconsistent buildup of transitions can be observed: Increasing
the pulse width of an input, that only induced a pulse on $S_1$, by \SI{1}{\fs},
caused a pulse propagation all the way up to $S_4$. This is a direct consequence
of the fact, that \reg\ suppresses pulse widths below a certain threshold, as
was shown in \cref{fig:delay_comparison}. For down-pulses on $A_0$ \reg\ even
delivers very nonphysical results as signal $S_0$ only starts to change after
every other signal had been triggered. We retraced this to an unfortunate series
of delays causing the signal closest to the input switch last, which is the
actual opposite of what is seen in analog simulations.  Finally note the
constant shifts in pulse widths, i.e., once a pulse appears on a signal its
pulse width differs from the input pulse solely by a constant additive
value. These values are very similar for each output such that very similar
output traces are achieved.

A smooth increase of pulse widths is naturally much better modeled by the
\gls{idm}. In the simulations we even observed a strict causality among $S_0$ to
$S_4$, i.e, $S_i$ showed a transition after all $S_j, j<i$ had also
switched. Compared to \reg\ this is a big improvement. Compared to \spice,
however, some inaccuracies are still observable. For example the quick increase
on $S_4$ in relation to $S_3$ is not well depicted. Possible causes are
inaccurate delay values extracted from the design (as reported in
\cite{OMFS20:INTEGRATION}) or the still nonoptimal description of multi-input
gates. Nonetheless, due to its accurate pulse width degradation coverage, the
\gls{idm} is able to provide overall realistic results.

\begin{table}[t]
  \centering
  \caption{Simulation time mean and variance $\sigma$ of the \adder.}
  \label{tab:runtime_adder}
  \begin{tabular}{c | r  r | r  r | c }
    & \multicolumn{2}{c|}{\reg} & \multicolumn{2}{c|}{\gls{idm}} & \multicolumn{1}{c}{} \\
    \# & $\overline{x}$ [\si{\s}] & $\sigma$ [\si{\s}] & $\overline{y}$ [\si{\s}] & $\sigma$ [\si{\s}] &
   overhead [\%] \\ \hline
    $1$ & $4.80$ & $0.92$ & $8.65$ & $0.90$ & $80.23$ \\ 
    $2$ & $5.95$ & $2.03$ & $12.00$ & $0.41$ & $101.58$ \\ 
    $4$ & $6.78$ & $0.90$ & $18.80$ & $0.86$ & $177.16$ \\ 
    $10$ & $11.74$ & $0.24$ & $37.75$ & $1.15$ & $221.43$ \\ 
    $20$ & $20.02$ & $0.42$ & $69.24$ & $2.09$ & $245.93$ \\ 
    $40$ & $37.30$ & $1.15$ & $132.53$ & $1.31$ & $255.27$ \\ 
    $100$ & $91.13$ & $2.19$ & $419.47$ & $105.57$ & $360.33$ \\ 
    $200$ & $216.17$ & $59.28$ & $1492.03$ & $317.88$ & $590.20$ \\ 
    $400$ & $1098.69$ & $242.03$ & $3674.48$ & $584.66$ & $234.44$ 
  \end{tabular}
\end{table}

\subsection{Overhead}

Calculating delay values for the \gls{idm}, which includes exponential and
logarithmic operations, is obviously computationally more expensive than
applying constant values paired with some minor removal checks for \reg. To
evaluate the overhead we ran extensive simulations and measured the execution
time (Intel Xeon X5650, 1600 MHz, 32 GB RAM, CentOS 6.10). As test circuits we
chose to use the \adder\ and the \clk\ of an open source MIPS
processor~\cite{Encounter_tutorial2016} that comprises of $227$ inverters which
drive $123$ flip-flops. To also generate results for larger circuits we simply
instantiated each unit multiple times.

For comparable results we had to ensure that \reg\ and the \gls{idm} process the
same amount of transitions. Since their behavior mainly differs for high input
frequencies we used rather long pulses to assure no internal cancellations,
whereat overall \num{2e5} input transitions were applied per simulation run. The
results are shown in \cref{tab:runtime_adder} for the \adder\ and in
\cref{tab:runtime_clk} for the \clk, whereat the first column denotes how often
the circuit had been instantiated. First and foremost we want to note that due
to a rather high variance $\sigma$ we ran each simulation $30$ times and
calculated the average $\overline{x}$ respectively $\overline{y}$. Furthermore
be advised that the presented values serve as lower bound, since real input
signals may lead to very short internal pulses which increases the workload of
\gls{idm} compared to \reg.

In essence the results show that the improved coverage of the \gls{idm}
definitely comes at a price.  For the \adder\ the overhead increases with
circuit size, while for $40$ instances it is almost $260$ \%. The significant
elevated values for $100$ and $200$ instances show a bottleneck of the
computational platform that is not experienced by both methods in the same
fashion, making them not representative. For the \clk\ the overhead is lower and
more constant, ranging from $27$ to almost $100$ \%. We explain the deviation by
the fact that only simple inverters are utilized, again showing that there is
still a lot to be done in the \gls{idm} regarding multi-input gates.

\begin{table}[t]
  \centering
  \caption{Simulation time mean and variance of the \clk.}
  \label{tab:runtime_clk}
  \begin{tabular}{c | r  r | r  r | c }
    & \multicolumn{2}{c|}{\reg} & \multicolumn{2}{c|}{\gls{idm}} & \multicolumn{1}{c}{} \\
    \# & $\overline{x}$ [\si{\s}] & $\sigma$ [\si{\s}] & $\overline{y}$ [\si{\s}] & $\sigma$ [\si{\s}] &
   overhead [\%] \\ \hline
    $1$ & $26.07$ & $2.18$ & $41.46$ & $1.12$ & $59.06$ \\ 
    $2$ & $41.17$ & $0.46$ & $69.58$ & $1.56$ & $69.01$ \\ 
    $4$ & $71.32$ & $1.27$ & $122.09$ & $1.25$ & $71.17$ \\ 
    $10$ & $188.27$ & $49.26$ & $368.30$ & $127.09$ & $95.62$ \\ 
    $20$ & $1016.23$ & $265.44$ & $1294.92$ & $451.77$ & $27.42$ \\ 
    $40$ & $2430.30$ & $406.60$ & $3554.59$ & $576.95$ & $46.26$ 
  \end{tabular}
\end{table}

\subsection{Summary}

Our simulation results have shown that \reg\ fails to model wide ranges of
undesired behaviors in the form of high frequency oscillations or metastable
intermediate voltages. The causes are single gates with larger delays, which
have to be expected in almost every real world circuit. Relying exclusively on
these predictions thus leads to a false sense of security. Investing sometimes
considerably more computational effort by applying the \gls{idm} leads to a much
better behavioral coverage and, in consequence, more trustworthy results. We
therefore claim that the \glsentrylong{idm} is able to enhance digital simulations
significantly.


\section{Conclusion and Future Work}
\label{sec:conclusion}

In this paper we evaluated the \glsentryfull{idm}, an elaborate alternative to
classic digital timing analysis approaches. To motivate this statement we ran
analog (\spice) and digital (inertial delay, \gls{idm}) simulations on three
different circuits (simple \boolOR-loop, an SR-latch and an adder) and compared
the derived results. Appropriate interpretation of the predictions by the
\gls{idm} confirmed the high behavioral coverage, especially for short
pulses. On the contrary a single high delay gate, which blocks a large share of
incoming pulses, caused massive mispredictions for inertial delay. Consequently
state-of-the-art simulation suites tend to miss potentially malicious circuit
states like infinite oscillations or metastability. Although an evaluation of
the overhead introduced by the \gls{idm} showed a significant increase in
simulation time, we still think that the \gls{idm} poses a viable alternative to
existing approaches, especially if confined to the most critical parts.

In the paper we argued, that a proper interpretation of the digital results,
i.e., whether a signal shows malicious behavior, is only possible by considering
the internal analog representation. Thus, future work will be devoted to
developing an extension that allows the designer to take a look at the
underlying analog traces on top of the digital results. Furthermore, we are
considering ways to improve the coverage of metastable behavior that is
currently not detectable at the digital output and of course performance
improvements.

One point deliberately neglected in this paper is accuracy. Investigations by
\"Ohlinger \textit{et al.}~\cite{OMFS20:INTEGRATION} and Maier \textit{et
  al.}~\cite{MFNS19:ASYNC} revealed that deriving appropriate description is a
nontrivial task. Furthermore, characterizing each single gate by relying heavily
on analog simulations is computationally expensive.  Approaches that yield
reasonable results based on available, or easily achievable, data are
instrumental for making the \gls{idm} a truly competitive alternative to
existing delay models.



\end{document}